\documentclass[review]{elsarticle}

\usepackage{graphicx}

\usepackage{amssymb}

\usepackage{lineno}

\begin{document}
\linenumbers

\begin{frontmatter}



\title{Development of the analog ASIC for multi-channel readout X-ray CCD camera}


 \author[label1]{Hiroshi Nakajima\corref{cor1}}
 \author[label2]{Daisuke Matsuura}
 \author[label2]{Toshihiro Idehara}
 \author[label1]{Naohisa Anabuki}
 \author[label1]{Hiroshi Tsunemi}
 \author[label3]{John P. Doty}
 \author[label4]{Hirokazu Ikeda}
 \author[label5]{Haruyoshi Katayama}
 \author[label6]{Hisashi Kitamura}
 \author[label6]{and Yukio Uchihori}
 \address[label1]{Department of Earth and Space Science, Graduate School of Science, Osaka
University, 1-1 Machikaneyama, Toyonaka, Osaka, 560-0043, Japan}
 \address[label2]{Department of Space and Integrated Defense Systems, Mitsubishi
Heavy Industries Ltd., 1200 Higashi Tanaka, Komaki, Aichi, 485-8561, Japan}
 \address[label3]{Noqsi Aerospace Ltd.,
2822 South Nova Road, Pine, Colorado 80470, USA}
 \address[label4]{the Institute of Space and Astronautical Science,
Japan Aerospace Exploration Agency,
3-1-1 Yoshinodai, Chuo-ku, Sagamihara, Kanagawa, 252-5210, Japan}
 \address[label5]{the Earth Observation Research Center,
Japan Aerospace Exploration Agency,
Tsukuba Space Center, 2-1-1 Sengen, Tsukuba, Ibaraki, 305-8505, Japan}
 \address[label6]{National Institute for Radiological Sciences (NIRS),
Anagawa 4-9-1, Inage-ku, Chiba-shi, Chiba, 263-8555, Japan}

\cortext[cor1]{Further author information: (Send correspondence to H.N.)\\H.N.:
E-mail: nakajima@ess.sci.osaka-u.ac.jp, Telephone: +81 6 6850 5478}

\begin{abstract}
 We report on the performance of an analog application-specific
 integrated circuit (ASIC) developed aiming for the front-end electronics of
 the X-ray CCD camera system onboard the next X-ray astronomical satellite, ASTRO-H.
 It has four identical channels
 that simultaneously process the CCD signals. Distinctive capability
 of analog-to-digital conversion enables us to construct a CCD camera body
 that outputs only digital signals.
 As the result of the front-end electronics test, it works properly with low
 input noise of $\le$30$\mu$V at the pixel rate below 100~$\!$kHz.
 The power consumption is sufficiently low of $\sim$150~$\!$mW/chip.
 The input signal range of $\pm$20~$\!$mV covers the effective energy range of
 the typical X-ray photon counting CCD (up to 20~$\!$keV). The integrated non-linearity
 is 0.2~\% that is similar as those of the conventional CCDs in orbit.
 We also performed a radiation tolerance test against the total ionizing dose (TID) effect
 and the single event effect. The irradiation test using $^{60}$Co and
 proton beam showed that the ASIC has the sufficient tolerance against TID up to
 200~$\!$krad, which absolutely exceeds the expected amount of dose
 during the period of operating in a low-inclination low-earth orbit.
 The irradiation of Fe ions with the fluence of 5.2$\times$10$^8$~$\!$Ion/cm$^2$ resulted in
 no single event latchup (SEL), although there were some possible single event upsets.
 The threshold against SEL is higher than 1.68~MeV~$\!$cm$^2$/mg, which is sufficiently
 high enough that the SEL event should not be one of major causes of instrument downtime in orbit.
\end{abstract}

\begin{keyword}
X-ray \sep ASIC \sep CCD camera
\PACS 
\end{keyword}
\end{frontmatter}

\section{Introduction}
\label{sec:intro}
%
%
X-ray CCD (charge-coupled device) camera has achieved primary roles
in X-ray astronomy thanks to its well-balanced performances:
a moderate energy resolution of $\sim$130~eV (FWHM) @ 5.9~keV
\citep{Short98,Garmire03,Koyama07}, a satisfactorily high quantum efficiency
in the wide energy range from 0.3 to 12~keV, and a high positional resolution
of $\sim$10~$\mu$m. Conventional frame-transfer CCDs
equip one readout node per several millions of pixels. To process the
frame image in several seconds, the readout electronics tend to
be very large and power-dissipated.

Recently some payloads in astronomical satellites adopt
application-specific integrated circuit (ASIC) to achieve extremely
lower power consumption and smaller size of the electronics than
those made of discrete ICs \citep{Rando03,Tajima04,Herrmann07}.
Meidinger et al. \citep{Meidinger06} reported that their CCD camera
system for a future astronomical mission with a thick (450~$\mu$m),
fully depleted pnCCD and ASIC chip proved the low readout noise of
2~e$^-$ rms and fast pixel rate of 13~MHz. All of the above ASICs,
however, only manipulate analog signals or barely have a discriminator.
To suppress the potential readout noise more strictly, therefore, we
have been developing an analog complementary Metal-Oxide-Semiconductor
(CMOS) ASIC with an analog-to-digital (AD) conversion
capability and placed it next to the CCD. We employed $\Delta\Sigma$
modulators \cite{Inose62} as the AD converter (ADC) since this type
of ADC can achieve a high resolution at a moderately short
conversion time and functions not only as an ADC but also a band-pass filter.

The main objective of our ASIC is the ASTRO-H satellite \citep{Takahashi10},
Japanese X-ray astronomical satellite that will be launched in 2014.
The X-ray CCD camera (SXI: Soft X-ray Imager) \citep{Tsunemi10} takes a role
of the primary X-ray imager.
It consists of four CCD chips that are abutted into a 2 by 2 array. Each chip
has an imaging area of 1280 $\times$ 1280 pixels and four readout nodes. Given
the focal length of 5.6~$\!$m, the four chips cover a region of 38~$\!^\prime\times$ 38~$\!^\prime$
on the sky combined with the Soft X-ray Telescope \citep{Awaki10}. SXI will utilize
back-illuminated (BI) chips, with a thick depletion layer of $>$200~$\!$$\mu$m.
Then effective energy range will be 0.4 - 12~$\!$keV, which bridges the ranges of
other instruments, the Soft X-ray Spectrometer (0.3 - 10~$\!$keV) \cite{Mitsuda10}
and the Hard X-ray Imager (5 - 80~$\!$keV) \cite{Kokubun10}.

As well as the basic functions and performances as the front-end electronics of SXI,
the radiation tolerance must be investigated to verify the functionality
on the planned orbit throughout the mission lifetime.
The radiation damage to the MOS Field-Effect Transistors (MOSFETs)
causes a total ionizing dose (TID) effect and a single event
effect (SEE). As a result of the TID effect, which is due to the
protons and electrons, the leak current increases, the threshold voltages
of MOSFETs change, and the 1/$f$ noise increases \citep{Fleetwood02}. The SEE, caused
by heavy-ions in cosmic-rays, produces many
hole-electron pairs in a specific MOSFET, hence causes non-destructive
SEE such as single event upset (SEU) or destructive SEE such as single event latchup
(SEL). Gamma-rays and particle beams are usually used to evaluate
the threshold level against the TID effect and the SEE.

Followed by the outline of the ASIC in Section \ref{sect:asic_spec}, we will
describe the results of the front-end electronics tests (Section \ref{sect:results}),
the radiation tolerance tests (Section \ref{sect:radtor}), and the summary
(Section \ref{sect:summary}). Indicated errors below mean 90~\% confidence level,
unless otherwise mentioned.

\section{Description of the ASIC}
\label{sect:asic_spec}

\begin{figure*}[ht]
 \begin{center}
  \begin{tabular}{c}
   \includegraphics[height=\textwidth, angle=-90]{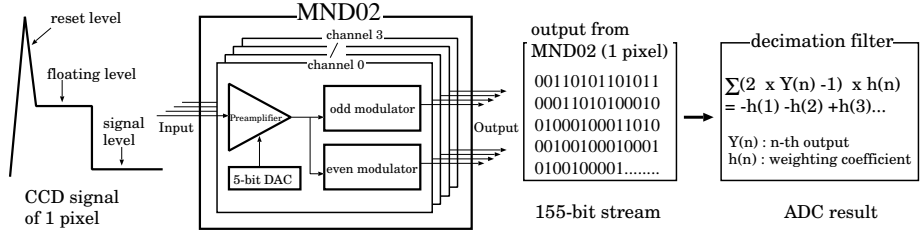}
  \end{tabular}
 \end{center}
 \caption[example] 
   { \label{fig:ASIC_howtoprocess} 
   The diagram of the signal processing of MND02. Input signals from CCD are at first
   amplified with adjustable gain.
   Simultaneously we put specific offset to the signal level in order to make
   use of dynamic range. Finally $\Delta\Sigma$ modulators
   convert analog signal of one pixel to digital 155-bit stream.
   Two modulators (``even" and ``odd") work alternately to improve the readout speed.
}
\end{figure*} 


Basic circuit configuration of our ASIC (hereafter we call it MND02) is
the same as that of MD01\citep{Matsuura07}. Here we summarize its specification
that characterizes our CCD camera system.

It was implemented through Taiwan Semiconductor Manufacturing Company (TSMC)
0.35~$\mu$m CMOS process and then the 3~mm square bare chip is packed into 15~mm square
ceramic quad flat pack.
The signal process in MND02 is shown in figure~\ref{fig:ASIC_howtoprocess}.
It works with 3.3~V power supply for analog and digital circuits.
Four identical circuits (from channel~$\!$0 to 3) process the CCD signals
simultaneously. Each circuit consists of a preamplifier, a 5-bit digital-to-analog
converter (DAC), and two $\Delta\Sigma$ modulators. Input signals are at first
amplified with adjustable gain from 0.6 to 10 in 9 steps.
The signal from CCD consists of reset, floating and signal levels. The voltage difference
between the latter two levels depends on the amount of electrons transferred by CCD.
However, there is a difference between these levels even in the case of no signal charge
due to the readout clocking pulse, which results as an offset in the `energy$-$pulse height'
relation. To minimize this offset and to effectively use MND02's input signal range,
we put an offset voltage to the signal level by 5-bit DAC.
Then $\Delta\Sigma$ modulators \citep{Doty06} take oversampling
for these levels and integrate them.
We measure the voltage difference by integrating two levels with opposite polarity.
This multiple sampling by the $\Delta\Sigma$ modulators
shifts the majority of the quantization noise above the signal band
in the frequency spectrum.
Finally it converts the analog signal of one pixel to a digital
155-bit stream. Two (``even" and ``odd") modulators work alternately to improve the
readout speed.


The bit stream is decoded by the decimation filter that is implemented in the
Field Programmable Gate Array to obtain a 12-bit decimal value.
We have determined the weighting coefficients for each bit in the filter
by simulations of our circuits (figure~6 in \citep{Matsuura07}) to
upgrade the frequency response as a low-pass filter and improve the signal-to-noise
ratio.


\section{Performance as a Front-end Electronics}
\label{sect:results}

We describe the results of the front-end electronics tests of MND02.
Pseudo CCD signals with constant input voltage were input for all of four channels
simultaneously. We tested 18 voltage levels throughout the input signal range
from --20 to +20~$\!$mV and obtained 819 pixel data from each voltage level.
Gain of the preamplifier was always set to be 10.
Readout rates were 1.25~$\!$MHz and its divisors of 2 until 19.5~$\!$kHz, which
is determined by the quartz crystal unit on the test module.

\begin{figure}
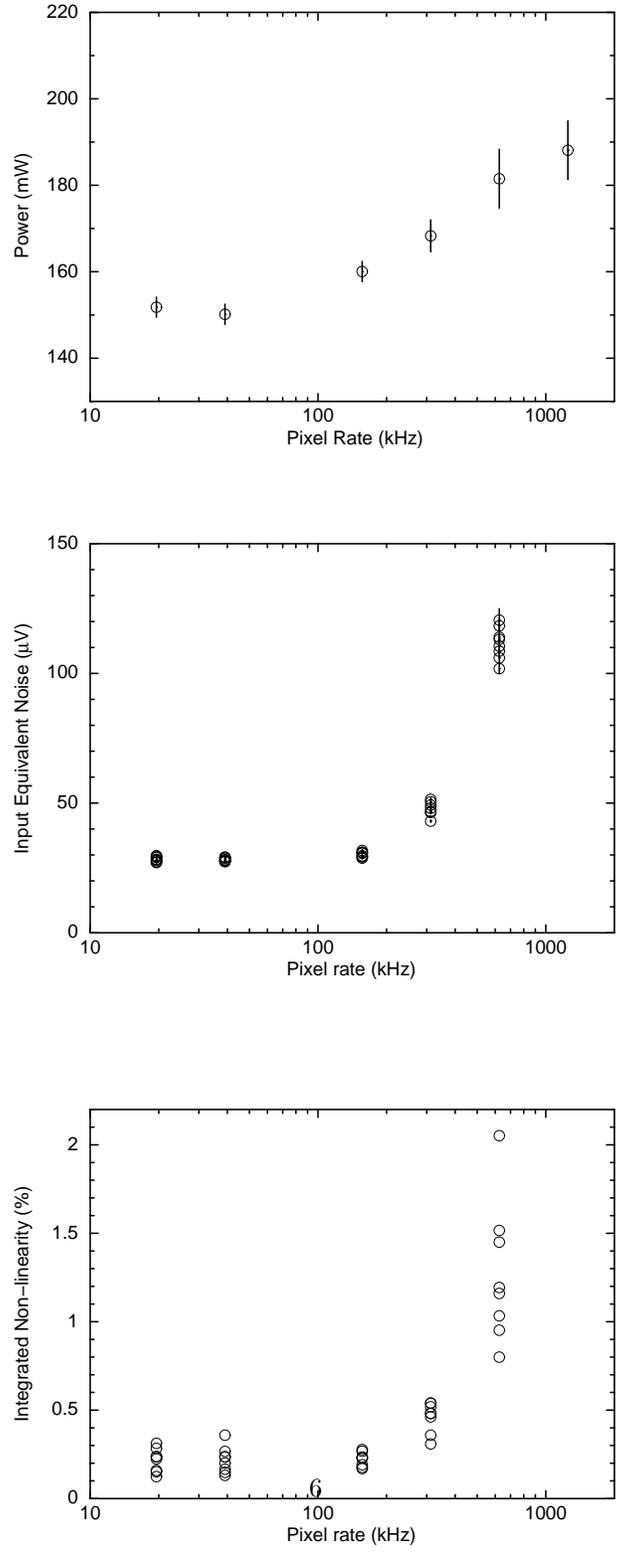

 \begin{center}
 \begin{tabular}{c}
   \includegraphics[height=8cm,angle=-90]{power_pixelrate_MND02.ps}
  \\
  \\
   \includegraphics[height=8cm,angle=-90]{noise_pixelrate_MND02_chall.ps}
  \\
  \\
  \\
   \includegraphics[height=8cm,angle=-90]{INL_pixelrate_MND02_chall.ps}
\end{tabular}
 \end{center}
   \caption[example] 
   { \label{fig:asic_unit}
Three parameters are shown as a function of pixel rate. Top panel:Power consumption, 
Middle panel:Input equivalent noise of the eight modulators, and
Bottom panel:The integrated non-linearity of the eight modulators.
}
 \end{figure}

\begin{figure}
 \begin{center}
  \begin{tabular}{c}
   \includegraphics[height=9.5cm,angle=-90]{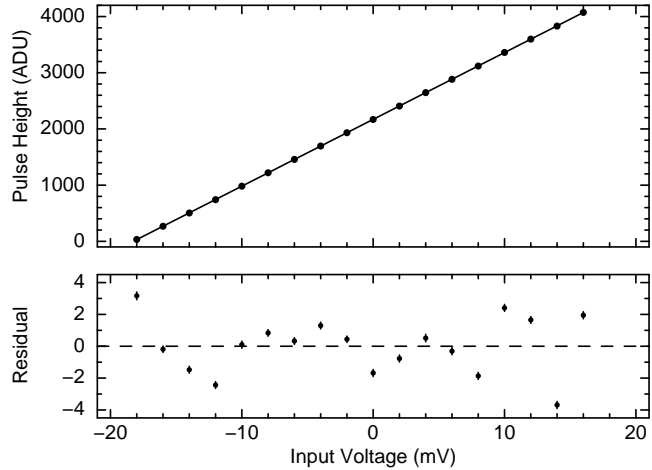}
  \end{tabular}
 \end{center}
 \caption[example] 
   { \label{fig:linearity} 
Top panel:Linearity plot throughout the input signal range of MND02 (channel~$\!$0,
even modulator) measured at the pixel rate of 78~$\!$kHz, the nearest value
to the pixel rate of SXI (68~$\!$kHz).
Bottom panel:Residuals from the best-fit linear function.
The INL is derived as 0.2\%.
}
\end{figure}

The power consumption was measured by comparing the current
in the printed circuit board (PCB)
between the case when MND02 was mounted and that when it was not on the PCB
(upper panel of figure~\ref{fig:asic_unit}).
Although the current increases with the readout rate, the power consumption
of the entire chip is about 150~$\!$mW at the pixel rate below 100~$\!$kHz.
In the middle panel of figure~\ref{fig:asic_unit}, we show the noise performance of MND02.
The statistical uncertainty of the decimal values was measured from
819 pixel data for each input voltage and we took the average of 18 data sets
taken with different voltage levels. We confirmed that the average noise was suppressed
to be less than 30~$\mu$V at the pixel rate below 100~$\!$kHz. 
Integrated non-linearity (INL) was also measured from the linearity plot
(figure~\ref{fig:linearity})
for each readout rate as shown in the bottom panel of figure~\ref{fig:asic_unit}.
Moderate INL of approximately 0.2\%
was obtained below 100~$\!$kHz.

\section{Radiation Tolerance}
\label{sect:radtor}

The first space application of MND02 will be
the ASTRO-H satellite that will be put into a low-earth orbit (LEO)
with a height of 550~$\!$km and an inclination angle of $<$30~$\!^{\circ}$.
The primary causes of the integrating damage in the LEO are protons and
electrons that penetrate a package when the satellite passes through the
South Atlantic Anomaly. We calculated the expected dose rate of Silicon assuming
the thickness of the Aluminum cover of 1~$\!$mm using the dose model of
SHIELDOSE-2\footnote{http://see.msfc.nasa.gov/ire/models.htm}.
AP8 and AE8 models are employed for the proton and electron spectra, respectively.
The dose rate obtained is 1.1~$\!$krad/year in this orbit.
Considering that the typical mission lifetime is $\sim$10~$\!$years and that
we should count the twice of margin, we require a tolerance of
$\ge$22~$\!$krad against the TID effect.

\subsection{Gamma-ray Test}
\label{ssec:cotest}

We adopted the $^{60}$Co (1.1 and 1.3~$\!$MeV) gamma-ray irradiation facility
at Radiation Research Center, Osaka Prefecture University
on 1 - 2 June 2009 for the TID effect test. No CCD 
was connected to the PCB and alternatively pseudo CCD signals were input at
a pixel rate of 78~$\!$kHz in the same manner as in the front-end electronics test.
The cylindrical $^{60}$Co with a height of 30~$\!$cm, 10~$\!$mm$\phi$,
an intensity of 186~$\!$TBq, was pulled up from underground to start the irradiation.
The ICs on the PCB except MND02 were protected by the lead with a thickness of
10~$\!$cm to suppress the gamma-ray intensity to be $<$10$^{-3}$ of that 
for MND02. The distance from the source to MND02 was determined such that
the intensity at the device under test was 28.6~$\!$krad/hour (200~$\!$krad
in 7~hours). We performed the above signal processing for only one chip during
the irradiation, while other five chips were biased, clocked but not processed.
All of the test was performed in the room temperature ($\sim$20~$\!^{\circ}$C),
which is the expected nominal thermal environment for MND02.

\begin{figure}
 \begin{center}
  \begin{tabular}{c}
  \hspace{-5mm}
   \includegraphics[height=9cm,angle=-90]{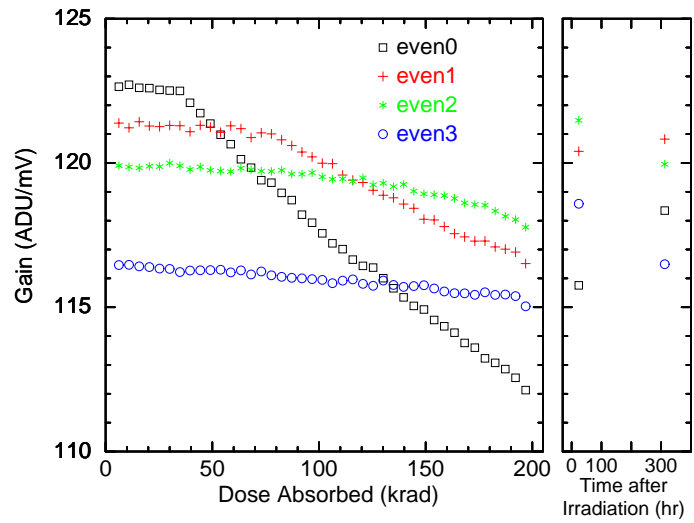}
  \end{tabular}
 \end{center}
 \caption[example] 
   { \label{fig:co60_gain} 
Left panel: Gain in the unit of ADU/mV as a function of the dose absorbed. Only the data of
even modulators are chosen for the brevity. We made the linearity plot and derived the gain
from the fitting results of the data with a linear function.
Right panel: Data obtained after the annealing are shown as a function of the time
after the irradiation test.
}
\end{figure} 

\begin{figure}
 \begin{center}
  \begin{tabular}{c}
  \hspace{-5mm}
   \includegraphics[height=9cm,angle=-90]{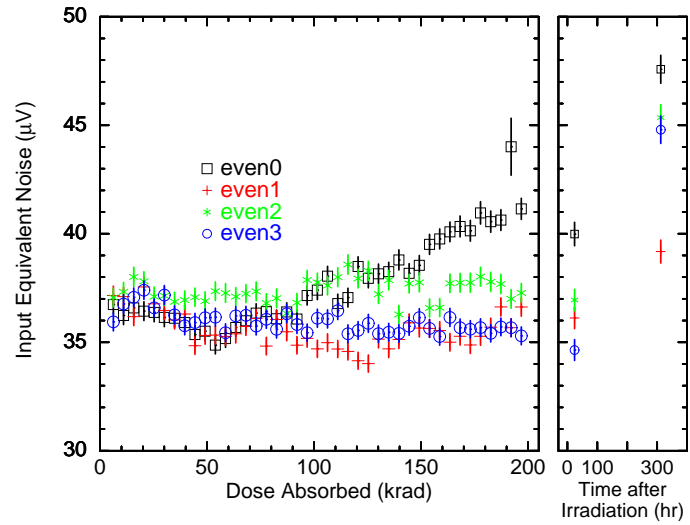}
  \end{tabular}
 \end{center}
 \caption[example] 
   { \label{fig:co60_noise} 
Left panel: Same as figure~\ref{fig:co60_gain}, but for the equivalent input noise.
We measured the dispersion of the decimal pulse heights as we did in
the front-end electronics test.
Right panel: The data obtained after the annealing are shown as a function of the time
after the irradiation test.
}
\end{figure} 

The current in MND02 increased by 10\% during the test with a total dose
of 200~$\!$krad.
Figure~\ref{fig:co60_gain} shows the gain variability as a function of the dose absorbed.
One of the modulator, even0 exhibits the earliest and most remarkable
gain degradation above $\sim$30~$\!$krad. Since a common behavior was seen
between even and odd modulators in the specific channel and the degradation was
gradual as a function of the dose absorbed, we think that the TID effect occurred
in the preamplifier rather than the modulator. All the other five chips exhibited the same
amount of degradation, and channel~$\!$0 in each chip showed the most remarkable degradation.
Since four channels are identical in electric circuit point of view,
the result suggests that there is an intrinsic weak point in the layout design of the
bare chip. Nevertheless, the degree of the degradation is less than 10\% even above
200~$\!$krad that corresponds to about 200~$\!$years in the LEO. Hence it will
not cause the practical problem in orbit.
Figure~\ref{fig:co60_noise} also guarantees the stable noise performance of all
the modulators during the mission lifetime expected.

The above performances were monitored also after the irradiating test
to investigate the degree of the annealing effect in the room temperature.
The right panels in figure~\ref{fig:co60_gain} and~\ref{fig:co60_noise} show the
performance variability as a function of the time after the irradiation.
The gain recovered in the post-irradiation test performed about two weeks after
although the channel~$\!$0 seemed to be still in recovery progress. The transition
of the input equivalent noise is a puzzle; all the six chips showed strong degradation
after 300 hours. However, we performed the front-end
electronics test again for the six chips about 16 months after the irradiation
and found that the input equivalent noise recovered to the same level as those
of non-irradiated chips. Hence we can say there is no permanent TID effect.

\subsection{Proton/Fe ion test}
\label{ssec:pfetest}

To deal with the proton/Fe ion beam test, we utilized the synchrotron ring at
Heavy Ion Medical Accelerator in Chiba (HIMAC)
in National Institute of Radiological Sciences (NIRS) on 6 - 8 July 2009.
A 150~$\!$MeV proton and a 400~$\!$MeV/amu Fe ion beam with a size of
$\sim$2~$\!$mm FWHM were irradiated to MND02. These energy values were
chosen since the irradiation facility for physics in HIMAC has many achievements
for the value. Since the bare chip is 3~$\!$mm square,
the efficiency that the particle hit the chip was calculated considering the
two-dimensional profile of the beam. All of the dose and fluence value hereafter
are calculated considering the efficiency.
Table~\ref{tab:beam_spec} summarizes the specification of both beams.

\begin{table}[h]
\caption{The specification of the beam at HIMAC.} 
\label{tab:beam_spec}
\begin{center}       
\begin{tabular}{|l|c|c|}
\hline
\rule[-1ex]{0pt}{3.5ex}  Species & Proton & Fe ion \\
\hline
\rule[-1ex]{0pt}{3.5ex}  Energy (MeV/amu) & 150 & 400 \\
\hline
\rule[-1ex]{0pt}{3.5ex}  Max. intensity (Ion/sec/cm$^2$) & 6.2$\times$10$^8$ & 1.3$\times$10$^6$ \\
\hline
\rule[-1ex]{0pt}{3.5ex}  Linear energy transfer &  4.38$\times$10$^{-3}$ & 1.68 \\
\rule[-1ex]{0pt}{3.5ex}  (MeV cm$^2$/mg) &  & \\
\hline
\end{tabular}
\end{center}
\end{table} 

We were anxious about possible performance change of the ASIC just after being put
into the orbit. Hence we began the test with relatively low intensity of 2.9~$\!$krad/hour
until the total dose amounted to be 0.7~$\!$krad. Since we confirmed that there was
no performance change, we increased the beam intensity to 158~$\!$krad/hour to
measure the TID tolerance.
The total dose absorbed was 167~$\!$krad through our experiment. Both of the two chips
under the test showed no performance degradation nor increase of the current including the
post-irradiation test. This is inconsistent with the result we observed in the $^{60}$Co test.
We suspect that the estimation of the dose rate with $^{60}$Co
is relatively ambiguous compared with that with protons. We calibrated the dose rate
in the $^{60}$Co irradiation facility using an ionization chamber and we needed to convert
the exposure dose to the absorbed dose for silicon. There may be some uncertainty
in these calibration processes.

These results exhibit the significant improvement of the TID tolerance
compared with that of the previous version of the ASIC, MD01 \citep{Nakajima09}.
The fabrication process of MD01 is the same as that of MND02, although the former chip
was packaged in the plastic QFP. There is also no change for the experiment condition
from the previous one except that the energy of the proton was 200~$\!$MeV and that
the intensity was slightly higher than that of this work.
This leads us to attribute the lack of the TID tolerance of MD01 to the circuit design
and/or the layout design. The major difference in the circuit design between MND02 and
MD01 is in the preamplifier; MND02 equips P channel MOSFETs at the first amplifying
stage since the P channel MOSFETs generally exhibit lower 1/$f$ noise than that of the
N channel MOSFETs. The size of the first stage FET in MND02 was set to be larger than
that in MD01. Although these revisions were designed for lower noise performance, we
found the radiation tolerance improved.

We tested three chips for Fe ion beam. The total fluence of the former two chips were
1.5$\times$10$^7$~$\!$Ion/cm$^2$ while that for the last chip was
5.2$\times$10$^8$~$\!$Ion/cm$^2$.
We performed functional test in the same manner as in the $^{60}$~$\!$Co test and
monitored the current in the PCB for analog and digital circuits. These values
were 14 - 15~$\!$mA and 119 - 120~$\!$mA for analog and digital circuits
throughout the irradiation, respectively. We found no peculiar increase of the current,
that is, no SEL occurred in the test. Hence the cross section against the SEL
($\sigma_{\rm SEL}$) was estimated assuming an upper limit of three events
for the fluence to be $< 5.8\times10^{-9}$~$\!$cm$^2$/(Ion$\times$ASIC)
at the LET of 1.68~$\!$MeV~$\!$cm$^2$/mg at 95\% confidence level.
This LET is sufficiently high enough that SEL event
should not be one of major causes of instrument downtime even if the recovery requires
power cycle of the instrument.

\begin{figure}[!t]
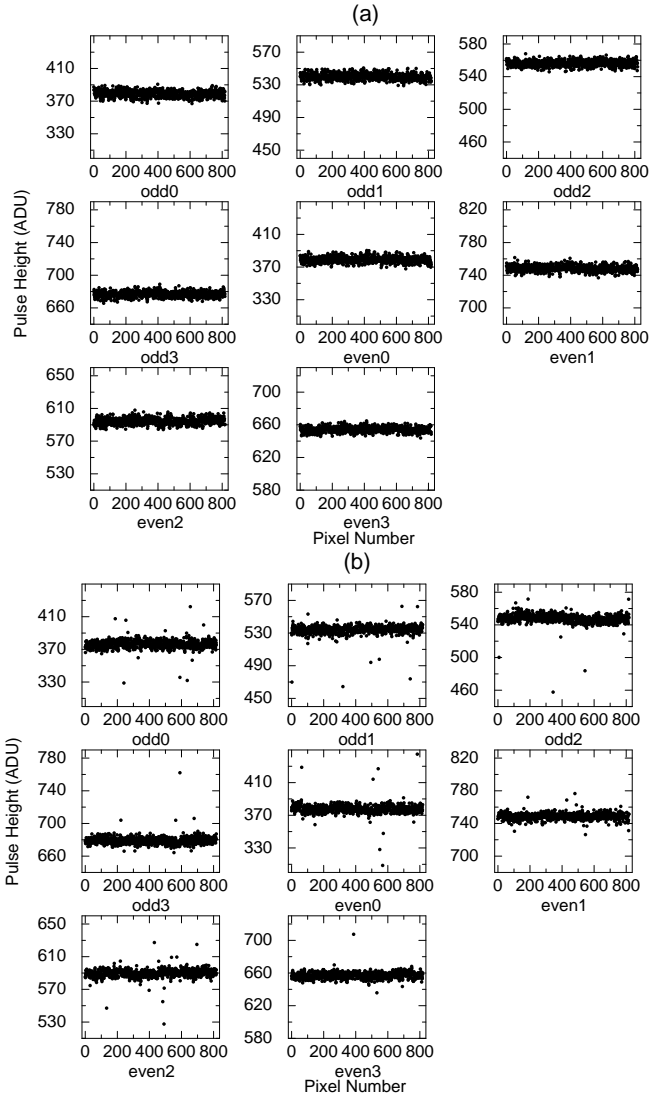

\vspace{3mm}
\hspace{-0mm}
\includegraphics[width=2.7in,angle=-90]{scatter_HIMACFe0krd_4paper_rev1.ps}
\vspace{4mm}
\\
\hspace{-0mm}
\includegraphics[width=2.7in,angle=-90]{scatter_HIMACFe3krd_4paper_rev1.ps}
 \caption[example] 
   { \label{fig:fe_scatter} 
Distribution of the decimal pulse height amplitudes when MND02
processed 819 pseudo CCD signals with an input voltage level of --18~$\!$mV
(a)~before the radiation tolerance test and (b)~during the irradiation.
Eight plots are shown for each $\Delta\Sigma$ modulator.
The vertical scales are selected so that we can see all the plots in (b).
}
\end{figure} 

Figure~\ref{fig:fe_scatter} shows the distribution of the decimal pulse height
for the data of 819~pixels with a constant input voltage. We observed some anomalous
pixels with the deviation of about 20 - 80~$\!$ADU from
the distribution center during the irradiation.
We calculated using the weighting coefficients of the decimation filter
that the typical magnitudes of the deviation should be about 120~$\!$ADU if we assume
a single-bit upset in the 155-bit stream. Hence the result suggests that
the upset occurred not in the output flip-flops. Then the suspected origin is that
some charges are injected into the integrators in the $\Delta\Sigma$
modulators or some capacitors in the preamplifiers.
The cross section of the SEU at the LET of 1.68~$\!$MeV~$\!$cm$^2$/mg was calculated
to be $\sigma_{\rm SEU} = 5.9\times10^{-9}$~$\!$cm$^2$/(Ion$\times$bit).

Future work to board our ASIC on ASTRO-H is to obtain the LET threshold
of $\sigma_{\rm SEU}$ and $\sigma_{\rm SEL}$ in order to estimate the SEE
rate in the LEO.
The burn-in test and the thermal functional test will also be performed.

\section{Summary} 
\label{sect:summary}

We summarize the results of the front-end electronics test
and the radiation tolerance test of MND02 as follows:

\begin{enumerate}

\item  As a result of the front-end electronics test, it works properly with low
 input noise of $\le$30$\mu$V at the pixel rate below 100~$\!$kHz.
 The power consumption is sufficiently low of $\sim$150~$\!$mW/chip.
 The INL is 0.2~\% throughout the input signal
 range from --20 to +20~$\!$mV, which is similar performance to the conventional
 electronics in orbit.

\item The radiation tolerance against the TID effect was estimated
by using $^{60}$Co and the 150~$\!$MeV proton beam.
The gain and input equivalent noise were stable
until the dose of 30~$\!$krad absorbed for the former test.
Considering the expected dose rate of 1.1~$\!$krad/yr in the LEO of ASTRO-H,
the radiation tolerance of MND02 indicates the proper operation for more than
20~years.

\item Fe ion beam test with the LET of 1.68~$\!$MeV~$\!$cm$^2$/mg
and the fluence up to 5.2$\times$10$^8$~$\!$Ion/cm$^2$
showed no SEL.
The threshold against the SEL is higher than 1.68~MeV~$\!$cm$^2$/mg,
which is sufficiently high enough that the SEL events should not be
one of major causes of the instrument downtime.

\end{enumerate}

We acknowledge Assistant Professor Takao Kojima of Osaka Prefecture University
and Hidenobu Mori who offered us much support and encouragement in the $^{60}$Co experiment.
This work is partly supported by the Nano-Satellite Research and Development Project in Japan
and the Research Project with Heavy Ions at NIRS-HIMAC.


\end{document}